\documentstyle[aps,multicol,psfig]{revtex}        

\begin{document}

\title{Lyapunov Potential Description for Laser Dynamics}

\author{Catalina Mayol$^{\dagger,\ddagger}$, Ra\'ul Toral$^{\dagger,\ddagger}$ 
and Claudio R. Mirasso$^{\dagger}$}

\address{($\dagger$) Departament de F\'{\i}sica, Universitat de les Illes Balears\\
($\ddagger$) Instituto Mediterr\'aneo 
de Estudios Avanzados (IMEDEA, UIB-CSIC)\\
E-07071 Palma de Mallorca, Spain. \\ Web Page: http://www.imedea.uib.es}  
\maketitle

\begin{abstract}

We describe the dynamical behavior of both class A and class B lasers
in terms of a Lyapunov potential.  For class A lasers we use the
potential to analyze both deterministic and stochastic dynamics. In
the stochastic case it is found that the phase of the electric field
drifts with time in the steady state. 
 
For class B lasers, the potential obtained is valid in the absence 
of noise. In this case,  a general expression relating the period of
the relaxation  oscillations to the potential is found. We have
included in this expression the terms corresponding to the gain
saturation and  the mean value of the spontaneously emitted power,
which were not considered previously. The validity of this
expression  is also discussed and a semi-empirical relation giving
the period of  the relaxation oscillations far from the stationary
state is proposed and checked against numerical simulations.
\end{abstract}
\vspace{1cm}

\noindent PACs: 42.65.Sf, 42.55.Ah, 42.60.Mi, 42.55.Px
\begin{multicols}{2}

\section{Introduction}
\label{introduccion}

Even for non-mechanical systems, it is occasionally possible to 
construct a function (called Lyapunov function or Lyapunov potential)
that decreases along trajectories \cite{lyapunov}. The usefulness of
Lyapunov functions lies on the fact that they allow an easy
determination of the fixed points of a dynamical (deterministic)
system as the extrema of the Lyapunov  function as well as
determining the stability of those fixed points. In some cases, the
existence of a Lyapunov potential allows an intuitive understanding
of the transient and stationary trajectories as movements of test
particles in the potential landscape. In the case of
non--deterministic dynamics, i.e. in the presence of noise terms, and
under some general conditions, the stationary probability
distribution can also be governed by the Lyapunov potential and
averages can be performed with respect to a known probability density
function. The aim of this work is to construct Lyapunov potentials
for some laser systems. We start, then, by briefly
reviewing the main features of the laser as a dynamical system.

A laser has three basic ingredients: i) a gain medium capable of
amplifying the electromagnetic  radiation propagating inside the
cavity,  ii) an optical cavity that provides the necessary feedback, 
and iii) a pumping mechanism.  A complete understanding of laser
dynamics is based on a fully quantum-mechanical description of 
matter-radiation interaction within the laser cavity. However, the
laser is a system where the number of photons  is much larger than
one, thus  allowing a semiclassical treatment of the electromagnetic
field inside the cavity through the Maxwell equations. This fact was
introduced in the  semiclassical laser theory, developed by
Lamb~\cite{lamb1,lamb2} and independently by
Haken~\cite{haken2,haken1,haken,syner}. This model for laser dynamics was
constructed from the Maxwell-Bloch equations  for a single-mode field
interacting with a two-level medium. The  semiclassical laser theory
ignores the quantum-mechanical nature of the electromagnetic field,
and the  amplifying medium is modeled quantum-mechanically as a
collection of two-level atoms through the Bloch  equations. A simpler
description can be obtained by deriving rate equations for  the
temporal change of the electric field (or photons number) inside the
cavity and the population inversion (carriers number in the case of
semiconductor lasers) \cite{tang}. Rate equations, with stochastic
terms accounting for spontaneous emission noise, have been
extensively used for semiconductor lasers. 

Different types of lasers can be classified according to the decay
rate of the photons, carriers and material polarization. Arecchi et
al.~\cite{arecchi} were the first to use a classification scheme: 
class C lasers have all the decay rates of the same order, and
therefore a set of three nonlinear differential equations is required
for a satisfactory description of the electric field, the population
inversion and the material polarization. For class B lasers, the
polarization decays towards the steady state much faster than the
other two variables, and it can be  adiabatically eliminated.  Class
B lasers, of which semiconductor lasers \cite{agrawal} are an
example, are then described by just two rate equations for the atomic
population  inversion (or carriers number) and the electric field.
Another examples of class B lasers are CO$_2$ lasers and  solid state
lasers \cite{vilaseca}. Finally, in class A lasers population
inversion and material polarization  decay much faster than the
electric field. Both material variables can be adiabatically 
eliminated, and the equation for the electric field is enough  to
describe the dynamical evolution of the system. Some properties of
class A lasers, like a dye laser, are studied in   \cite{ciu,her}. 
In this paper we interpret the dynamics of both class A
and class B  lasers by using a Lyapunov potential. 

The paper is organized as follows. In Section \ref{review} we present
a  brief review of the relation of Lyapunov potentials to the
dynamical equations and the splitting of those into conservative and
dissipative parts. We consider the example of class A lasers. In this
case, the Lyapunov potential gives an intuitive understanding of the
dynamics observed in the numerical simulations. In the presence of
noise,  the probability density function obtained from the potential
allows the  calculation of stationary mean values of interest as, for
example,  the mean value of the number of photons.  We will show that
the mean value of the phase of the electric field in the steady state
varies linearly with time only when noise is present, in a phenomenon
reminiscent of the noise--sustained flows. In Section \ref{classB},
the dynamics of rate equations for class B lasers is presented in
terms of the intensity  and the carriers number (we will restrict
ourselves to the semiconductor laser).  In this case we have found a
potential which helps to analyze the corresponding dynamics in the
absence of noise. By using the conservative part of the equations,
one can obtain an expression for the period of the oscillations in the
transient regime following the laser switch-on.  
This expression extends the one obtained in a simpler
case by an identification of the laser dynamics with a Toda
oscillator in \cite{oppo}. Here, we have  added in the expression for
the period the corresponding  modifications for the gain saturation
term and spontaneous emission noise. Finally, in section
\ref{summary}, we summarize the main results obtained.

\section{Potentials and Lyapunov Functions: Class A Lasers}
\label{review}
The evolution of a system (dynamical flow) can be classified into 
different categories according to the relation of the Lyapunov
potential to the actual  equations of motion~\cite{mon,mon2}. We
first consider a deterministic dynamical flow in  which the real
variables $(x_1, \dots,x_N)\equiv {\bf x}$ satisfy the general
evolution equations:
\begin{equation}
\frac{d x_i}{d t} = f_i({\bf x}), \hspace{2.0cm} i=1,\dots,N
\end{equation}
In the so--called {\sl potential flow},  there exists a non--constant
function $V({\bf{x}})$ (the potential) in terms of which the above
equations can be written as:
\begin{eqnarray}
\frac{d x_i}{d t}=-\sum\limits_{j=1}^N S_{ij} \frac{\partial
V}{\partial x_j} + v_i
\label{norela1}
\end{eqnarray}
where $S({\bf{x}})$ is a symmetric and positive definite matrix,
and $v_i({\bf x})$ satisfy the {\sl orthogonality condition}:
\begin{equation}
\sum\limits_{i=1}^N  v_{i} \frac{\partial V}{\partial x_i} = 0 
\label{norela3}.
\end{equation}
A {\sl non-potential flow}, on the other hand, is one for which the
splitting (\ref{norela1}), satisfying (\ref{norela3}), admits only
the trivial solution $V({\bf{x}})=$ constant, 
$v_i({\bf{x}})=f_i({\bf{x}})$. 

Since the above (sufficient) conditions for a potential flow lead to
$dV/dt \le 0$, one concludes that $V({\bf{x}})$ (when it satisfies
the additional condition of being  bounded from below) is a Lyapunov
potential for the dynamical system. In this case, one can get an
intuitive understanding of the dynamics: the fixed points are given
by the extrema of  $V({\bf{x}})$ and the trajectories relax
asymptotically towards the surface of minima of $V({\bf{x}})$. This
decay is produced by the only effect of the terms containing the
matrix  $S$ in Eq. (\ref{norela1}), since the dynamics induced by
$v_i$ conserves the potential, and $v_i({\bf{x}})$ represents the
residual dynamics on this minima surface. A particular case of
potential flow is given when $v_i({\bf x})$ can also be derived from
the potential, namely:
\begin{equation}
\frac{d x_i}{d t}=-\sum\limits_{j=1}^N D_{ij} \frac{\partial 
V}{\partial x_j} 
\label{potflow}
\end{equation} 
where the matrix $D({\bf{x}})= S({\bf{x}})+ A({\bf{x}})$, 
splits into a positive definite symmetric matrix, $S$, and an
antisymmetric  one, $A$.  In this case, the residual dynamics also
ceases after the surface of minima of $V({\bf{x}})$ has been reached.

We now describe the effect of noise on the dynamics of the above systems. 
The stochastic equations (considered in the It\^{o} sense) are:
\begin{equation}
\label{eq:noise}
\frac{d x_i}{d t} = f_i({\bf x}) + \sum_{j=1}^N g_{ij}({\bf x}) \xi_j(t)
\end{equation}
where $g_{ij}({\bf x})$ are given functions and $\xi_j(t)$ are white
noise: Gaussian random processes of zero mean and correlations:
\begin{equation}
\label{noisecor}
\langle \xi_i(t)\xi_j(t')\rangle = 2 \epsilon \delta_{ij}\delta(t-t')
\end{equation}
$\epsilon$ is the intensity of the noise.

In the presence of noise terms, it is not adequate to talk about
fixed points of the dynamics, but rather consider instead the maxima
of  the probability density function $P({\bf{x}},t)$,  which
satisfies the multivariate Fokker-Planck equation \cite{risken,maxi}
whose general solution is unknown.  When the deterministic part of
(\ref{eq:noise}) is a potential flow,  however, a closed form for the
stationary distribution $P_{st}({\bf x})$ can be given in terms of
the potential $V({\bf x})$ if the  following (sufficient) conditions
are satisfied:
\begin{enumerate}
\item \label{cond1} The {\sl fluctuation--dissipation} condition,
relating the  symmetric matrix $S$ to the noise matrix $g$:
\begin{equation}
S_{ij} = \sum_{k=1}^N g_{ik} g_{jk}, \hspace{1.0cm} S= g\cdot g^T
\end{equation}
\item \label{cond2} $S_{ij}$ satisfies:
\begin{equation}
\sum_{j=1}^N \frac{\partial S_{ij}}{\partial x_j} = 0, 
\hspace{2.0cm} \forall i
\end{equation}
This condition is satisfied, for instance, for a constant matrix $S$.
\item \label{cond3} $v_i$ is divergence free:
\begin{equation}
\sum_{i=1}^N \frac{\partial v_i}{\partial x_i} = 0
\end{equation}
\end{enumerate}
this third condition is automatically satisfied for potential 
flows of the form (\ref{potflow}) with a constant matrix A. 

Under those circumstances, the stationary probability density
function is:
\begin{equation}
P_{st}({\bf{x}})=Z^{-1} \exp \left(-\frac {V({\bf{x}})}{\epsilon}\right)
\label{prob}
\end{equation}
where $Z$ is a normalization constant. 
Graham \cite{graham} has shown
that if conditions \ref{cond2} and \ref{cond3} are not  satisfied,
then the above expression for $P_{st}({\bf{x}})$ is still valid in
the limit $\epsilon \rightarrow 0$.

As an example of the use of Lyapunov potentials in a dynamical
system, we consider class A lasers \cite{haken} whose dynamics can be
described in terms of the slowly varying complex amplitude $E$ of
the electric field:
\begin{equation}
\label{eq11}
\dot{E}=(1+ i \alpha) \left(\frac{\Gamma } {1 + \beta |E|^2} - 
\kappa\right)~E + \zeta(t)
\end{equation}
where $\alpha$, $\beta$, $\Gamma$  and $\kappa$ are real parameters.
$\kappa$ is the cavity decay rate; $\Gamma$ the gain parameter; 
$\beta$ the saturation-intensity parameter and $\alpha$ is the
detuning  parameter. Another widely
used model expands the non-linear term to give a cubic  dependence on
the field (third order Lamb theory \cite{lamb1}), but this is not
necessary here. Eq. (\ref{eq11}) is written in a reference frame in which the frequency
of the {\sl  on} steady state is zero \cite{ciu}.
$\zeta(t)$ is
a complex Langevin source term accounting for the stochastic  nature
of spontaneous emission. It is taken as a Gaussian white noise of 
zero mean and correlations:
\begin{equation}
\langle \zeta (t) \zeta^{*} (t') \rangle = 4
\Delta \delta (t - t') ~~~~~~,
\end{equation}
where $\Delta$ measures the strength of the noise. 

By writing the complex variable $E$ as $E=x_1+i x_2$ and introducing
a  new dimensionless time such that $t \to \kappa t$, the evolution equations
become: 
\begin{eqnarray}
\label{x1an}
\dot x_1 & = & \left(\frac{a}{b+x_1^2+x_2^2}-1\right)~(x_1-\alpha x_2)+ \xi_1(t)\\
\label{x2an}
\dot x_2 & = & \left(\frac{a}{b+x_1^2+x_2^2}-1\right)~(\alpha x_1+x_2)+ \xi_2(t)
\end{eqnarray}
Where $a= \Gamma/(\kappa \beta)$ and $ b= 1/\beta$.
$\xi_1 (t)$ and $\xi_2(t)$ are white noise terms with zero
mean and correlations given by equation (\ref{noisecor}) with 
$\epsilon = \Delta /\kappa$.

In the deterministic case ($\epsilon=0$), these dynamical equations 
constitute a potential flow of the form
(\ref{potflow}) 
where the potential $V(\bf{x})$ is \cite{syner}
\begin {equation}
V(x_1,x_2)= \frac{1}{2}~ [x_1^2+x_2^2~-~a ~\ln
(b+x_1^2+x_2^2)]
\label{potclasa}
\end {equation}
and the matrix $D(\bf{x})$ (split into symmetric and antisymmetric
parts) is:
\begin {equation}
D=S+A=
\left(\begin{array}{cc}
                1 & 0 \\
                0 & 1
       \end {array} \right)                       
+ 
\left(\begin{array}{cc}
                0 & -\alpha \\
                \alpha & 0
       \end {array} \right)  ~~~~~~.     
\end{equation}                           
A simpler expression for the potential is given in  \cite{haken} and
\cite{risken} valid for the case in which the gain term is expanded in
Taylor series.

According to our discussion above, the fixed points of the
deterministic dynamics are the extrema of the potential $V({\bf x})$:
for $a>b$ there is a maximum at $(x_1,x_2)=0$   (corresponding to the
laser in the  {\sl off} state) and a line of minima given by
$x_1^2+x_2^2=a-b$  (see Fig. 1).  The asymptotic stable situation,
then, is that the laser switches to the {\sl on} state with an
intensity $I\equiv |E|^2 = x_1^2+x_2^2 = a-b$. For $a<b$ the only
stable fixed point is the  {\sl off} state $I=0$.

In the transient dynamics, the symmetric matrix $S$ is responsible
for driving the system towards the line of minima of $V$ following
the lines of maximum slope of $V$. The antisymmetric part $A$ (which
is proportional to $\alpha$) induces a movement orthogonal to the
direction of maximum variation of $V(\bf{x})$. The combined effects
of $S$ and $A$ produce a spiraling trajectory in the $(x_1,x_2)$
plane.  The angular velocity of this spiral movement is proportional
to $\alpha$. Asymptotically, the system tends to one of the minima in
the line $I=a-b$, the exact location depending on the initial
conditions. The potential decreases in time until it arrives at its
minimum value: $V(x_1^2+x_2^2=a-b)=-\frac{1}{2}(a ~\ln(a)-a
+b)$.        

In the presence of moderate levels of noise, $\epsilon \ne 0$, the
qualitative features of the transient dynamics remain the same as
in the deterministic case. The most important differences appear near
the stationary situation. As the final value of the intensity is
approached and for $\alpha \ne 0$,  the phase rotation slows down and
the mean value of the phase $\phi$ of the electric field $E$ changes
linearly with time also in  the steady state, see Fig. 2.  For
$\alpha=0$ there is only the ordinary phase diffusion around the
circumference $x_1^2+x_2^2=a-b$ that represents the set of all
possible deterministic equilibrium states \cite{ciu}. Therefore, for
$\alpha \ne 0$ the real and imaginary parts of $E$  oscillate not
only in the transient dynamics but also in the steady  state, and
while the frequency of the oscillations still depends on  $\alpha$
(as well as $\epsilon$),  their amplitude depends on the noise
strength $\epsilon$. 

We can understand these aforementioned features of the noisy
dynamics  using the deterministic Lyapunov potential $V(x_1,x_2)$.
Since conditions 1-3 above are satisfied, the stationary probability
distribution is given by (\ref{prob}) with $V(x_1,x_2)$ given by
(\ref{potclasa}). By changing variables to intensity and phase, we
find that the probability density functions for $I$ and $\phi$ are
independent functions,  $P_{st}(\phi)=1/(2\pi)$ is a constant and, 
\begin{equation}
P_{st}(I) = Z^{-1} {\rm e}^{-I/(2\epsilon)} ~(b+I)^{a/(2 \epsilon)}
\label {psti}
\end{equation}
where the normalization constant is
$Z=(2\epsilon)^{\frac{a}{2\epsilon}+1} 
{\rm e}^{\frac{b}{2\epsilon}}\Gamma\left(\frac{a}{2\epsilon}+1, 
\frac{b}{2\epsilon}\right)$ and $\Gamma(x,y)$ is the incomplete Gamma
function.  From this expression, we see  that, independently of the
value for $\epsilon$, $P_{st}(I)$ has its maxima at the deterministic
stationary value $I_{m}=a-b$. Starting from a given initial condition
corresponding, for instance, to the laser in the {\sl off} state, the
intensity fluctuates around a mean value that increases monotonically
with time. In the stationary state, the intensity fluctuates around
the deterministic value $I_m=a-b$ but, since the  distribution
(\ref{psti}) is not symmetric around $I_m$, the mean value $\langle I
\rangle_{st}$ is larger than the deterministic value.  By using
(\ref{psti}) one can easily find that
\begin{equation}
\label{meani}
\langle I \rangle_{st} =  (a-b)~+ 
~2 \epsilon ~\left[1 +\frac{\exp(-b/2\epsilon)
(b/2\epsilon)^{\frac{a}{2\epsilon}+1}}
{\Gamma\left(\frac{a}{2 \epsilon} +1, \frac{b}{2 \epsilon}\right)}\right]
\end{equation}
An expression for the mean value of the intensity in the steady state
was also  given in \cite{risken} in the simpler case of an expansion
of the saturation-term parameter in the dynamical equations.

As mentioned before, in the steady state of the stochastic dynamics,
the phase $\phi$ of the electric field fluctuates around a mean value
that changes  linearly with time. Of course, since any value of
$\phi$ can be mapped into the interval $[0,2\pi)$, this is not
inconsistent with the fact  that the stationary distribution for
$\phi$ is a uniform one. We can easily understand the origin of this 
{\sl noise sustained flow} \cite{magnasco}: the rotation inducing
terms, those proportional to $\alpha$ in the equations of motion, are
zero at the line of minima of the potential $V$ and, hence, do not
act in the steady deterministic state. Fluctuations allow the system
to explore regions of the configuration space $(x_1,x_2)$ where the
potential is not at its minimum value. Since, according to Eq.
(\ref{meani}), the mean value of $I$ is not at the minimum of the
potential, there is, on average, a non-zero contribution of the
rotation terms producing the phase drift observed.

The rotation speed can be calculated by writing the evolution 
equation for the phase of the electric field as:
\begin{equation}
\dot \phi = \left( \frac{a}{b+I} -1 \right) \alpha + \frac{1}
{\sqrt{I}}\xi(t)
\end{equation}
where $\xi(t)$ is a white noise term with zero mean value and
correlations  given by (\ref{noisecor}).
By taking the average value and using the rules of the It\^o
calculus, 
one arrives at:
\begin{equation}
\langle \dot \phi \rangle = \alpha \left\langle \frac{a}{b+I}-1\right\rangle
\label{dotmean}
\end{equation}
and, by using the distribution (\ref{psti}), one obtains the
stochastic frequency shift:
\begin{equation}
\langle \dot \phi \rangle_{st} = -\alpha \frac{\exp(-b/2 \epsilon)
(b / 2 \epsilon)^{\frac{a}{2 \epsilon}}}{\Gamma\left(
\frac{a}{2 \epsilon}+1, \frac{b}{2 \epsilon}\right)}
\label{fist}
\end{equation}
Notice that this average rotation speed is zero in the case of no
detuning ($\alpha=0$) or for  the deterministic dynamics ($\epsilon=0$)
and that, due to the minus sign, the rotation speed is opposite to
that of the deterministic transient dynamics when starting from the
{\sl off} state. These results are in excellent agreement with
numerical simulations of the  rate equations in the presence of noise
(see Fig. 2). 

\section{Class B lasers}
\label{classB}

The dynamics of a typical class B laser, for instance a single mode 
semiconductor laser, can be described in terms of two evolution
equations, one for the slowly-varying complex amplitude $E$ of  the
electric field inside the laser cavity and the other for the carriers number
$N$ (or electron-hole pairs) \cite{agrawal}. These equations include noise terms
accounting for the stochastic  nature of spontaneous emission and 
random non-radiative carrier
recombination due to thermal fluctuations. Both noise sources are usually
assumed to be white Gaussian noise.  

The equation for the electric field
can be written in terms of the optical intensity $I$ and the
phase $\phi$ by defining $E=\sqrt{I}\,e^{{\rm i} \phi}$. For 
simplicity, we neglect the explicit random fluctuations terms and 
retain, as usual \cite{agrawal}, the mean power of the spontaneous
emission.
The equations are:
\begin{equation}
\frac{dI}{dt}=(G(N,I) - \gamma) I + 4 \beta N 
\label{equai}
\end{equation}
\begin{equation}
\frac{d \phi}{dt}= \frac{1}{2} (G(N,I)-\gamma) \alpha 
\label{phase}
\end{equation}
\begin{equation}
\frac{dN}{dt}=C-\gamma_e N- G(N,I)I
\label{port}
\end{equation}
$G(N,I)$ is the material gain given by: 
\begin{equation}
G(N,I) = \frac{g(N-N_o)}{1+s I} 
\end{equation}
The definitions and typical values of the parameters for
semiconductor lasers are given in Table 1. 
The first term of Eq. (\ref{equai}) accounts for the stimulated
emission while the second accounts for the mean value of the
spontaneous emission power. Eqs. (\ref{equai} - \ref{port}) are
written in the reference frame in which  the frequency of the {\sl 
on} state is zero when spontaneous  emission noise is neglected.
The threshold condition is obtained by setting $G(N,I)=\gamma$, $I=0$
and neglecting spontaneous emission, i.e.
$N_{th}=N_o+\frac{\gamma}{g}$. The threshold carrier injected per
unit time to turn the laser {\sl  on} is given by $C_{th}=\gamma_e
N_{th}$. Eq. (\ref{phase}) shows that 
$\dot{\phi}$ is linear with $N$ and slightly (due to the
smallness of the saturation parameter $s$, see Table 1) nonlinear with $I$. 

Since in the deterministic case considered henceforth, the evolution
equations for $I$ and $N$  do not depend on the phase $\phi$, we can
concentrate only in the evolution of $I$ and $N$.  One can obtain a
set of simpler dimensionless equations by performing the  following
change of variables: 
\begin {equation}
y=\frac{2 g}{\gamma} I,
~~~
z=\frac{g}{\gamma}~(N-N_o),
~~~
\tau=\frac{\gamma}{2}t
\end{equation}

The equations become then:
\begin {eqnarray}
\frac{d y}{d\tau} & = &  2 \left(\frac{z}{1+\bar{s} y}-1\right)~y + 
c \, z + d \label{ibar} \\
\frac{d z}{d\tau} & = & a-b z- \frac{z~y}{1+\bar{s} y}~~~~,
\label{x}
\end{eqnarray}
where we have defined
$a=\frac{2 g }{\gamma^2}(C-\gamma_e N_o)$,
$b=\frac{2 \gamma_e}{\gamma}$, $c=\frac{16 \beta}{\gamma}$, $d=\frac{16 
\beta g N_o}{\gamma^2}$ and $\bar{s}=\frac{s\gamma}{2g}$.
These equations form the basis of our subsequent analysis. The steady
states are obtained by setting (\ref{ibar}) and (\ref{x}) equal to
zero, i.e.:
\begin{eqnarray}\label{yst}
y_{st} & = & \frac{1}{4 (1+b \bar{s})}[ 2 (a-b) +d (1+b \bar{s}) 
+c \, a \, \bar{s} + \sqrt{v}] \\
z_{st} & = & \frac{a (1 + \bar{s} y_{st})}{b+y_{st}(1+b \bar{s})}
\end{eqnarray}
where the constant $v$ is given by:
\begin{eqnarray}
v&=&4 (a-b)^2 + 4 d (a+b) (1+b \bar{s}) +d^2 (1+b \bar{s})^2 \nonumber \\
&+& c(8 a + 4 a \bar{s} (a +b) + 2 d a \bar{s} 
(1 +b \bar{s}))+c^2 a^2 \bar{s}^2
\end{eqnarray}
There is another steady state solution for $y_{st}$ given by Eq. 
(\ref{yst}) (with a minus sign
in front of $\sqrt{v}$) which, however, does not correspond to any
possible physical situation, since $y_{st}<0$. For a value of the
injected carriers per unit time below threshold ($C<C_{th}$,
equivalent to $a-b<0$) $y_{st}$ is very small. This corresponds to the
{\sl off} solution in which the only emitted light corresponds to the
spontaneous emission. 
Above threshold, stimulated emission occurs and the laser operates in
the {\sl on} state with large $y_{st}$. In what
follows, we will concentrate in the evolution following the laser
switch-on to the {\sl on} state. 

It is known that the dynamical evolution of $y$ and $z$ is such that
they both reach the steady state by performing damped 
oscillations~\cite{agrawal} whose period decreases with time. This
fact is different  from the usual relaxation oscillations that are
calculated near the steady state by linearizing the dynamical
equations.  The time evolution of $y$ and $z$ is shown in Fig. 3a for
some parameters (for another values of the parameters equivalent
results are obtained), while the corresponding projection in the $y$,
$z$ phase-plane is shown in  Fig. 4. We are interested in obtaining a
Lyapunov potential that can helps to explain the observed dynamics. 
This study was done in \cite{oppo} without considering neither the
saturation term, nor the mean value of the spontaneously emission 
power, and under those conditions an expression for the period of the
transient oscillations was obtained. In our work, we calculate the
period of the oscillations by taking into account these two effects. 
The period is obtained in terms of the potential, by assuming that
the latter has a constant value during one period. It will be shown that
this assumption works reasonably well and gives a good agreement with
numerical calculations. Near the steady state, the relaxation
oscillations can also be calculated in this form, but the potential
is almost constant and consequently so is the period.

The evolutions equations (\ref{ibar}), (\ref{x}) can be casted in the
form of Eq. (\ref{potflow}) with the following Lyapunov potential:
\begin{equation} 
V(y,z)=A_1~y 
+ A_2~y^2 + A_3 \ln(y) +
\frac{A_4}{y} 
+ \frac{1}{2} ~B^2(y,z)
\label{vcbs}
\end {equation}
where 
\begin{eqnarray} 
A_1 & = & \frac{1}{2}-\frac{1}{2} a \bar{s} + b \bar{s} - \frac{1}{4} 
\bar{s} d ~( 1+ b \bar{s}) -
\frac{1}{4} a \bar{s}^2 c \nonumber \\
A_2 & = & \frac{\bar{s}}{4}  ~( 1 + b \bar{s})  \nonumber \\ 
A_3 & = & -\frac{1}{2}~[a-b + (ac  + b d) ~ \bar{s} + \frac{d}{2}] \nonumber \\ 
A_4 & = & \frac{(ac + bd)}{4} \nonumber \\
B(y,z) & = & z-1-\bar{s} y + 
\frac{(d+c z)}{2 y} ~ (1+ \bar{s} y)~~~.\nonumber
\end{eqnarray}
The corresponding (non-constant) matrix $D$ is given by:
\begin{equation}
D=
\pmatrix{ 0 & -d_{12} \cr
d_{12} & d_{22} }.
\end{equation}
being
\begin{eqnarray}
d_{12} & = & \frac{4 y^2} {(1 +\bar{s} y) ~[2 y + c ~(1 + \bar{s} y)]} \\
d_{22} & = &\frac{4 y~[(1+2\bar{s}+ b \bar{s}) ~y^2 + b y + d + c z]} 
{(1+\bar{s} y)~[2 y + c~(1+ \bar{s} y)]^2}
\end{eqnarray}
 
This potential reduces to the one obtained in ref.~\cite{oppo} when 
setting $c=d=\bar{s}=0$ (which corresponds to set the laser
parameters $\beta=s=0$).  As expected, non-vanishing values for the
parameters  $s$ and $\beta$ increase the dissipative part of the
potential ($d_{22}$), associated with the damping term. This result
was pointed out in \cite{lee} when linearizing the rate equations
around the steady state.

The equipotential lines of (\ref{vcbs}) are also plotted in Fig. 4. 
It is observed that there is only one minimum for V and hence the 
only stable solution (for this
range of parameters) is that the laser switches to the {\sl on} state
and relaxes to the minimum of $V$. The movement towards the minimum
of $V$ has two components: a conservative one that produces closed
equipotential trajectories and a damping that decreases the value of
the potential. The combined effects drives the system to the minimum
following a spiral movement, best observed in Fig. 4.

The time evolution of the potential is also plotted in  Fig. 3b. In
this figure it can be seen  that the Lyapunov potential is
approximately constant between  two consecutive peaks of  the
relaxation oscillations (This fact can be also observed 
with the equipotential lines of Fig. 4). This fact allows us to estimate the
relaxation  oscillation period by approximating $V(y,z)= V$,
constant, during this time interval. When the potential is considered
to be constant, the period can be evaluated by the standard method of
elementary Mechanics: $z$ is replaced by its expression obtained from
(\ref{ibar}) in terms of $y$ and $\dot{y}$ (the dot stands for the
time derivative) in $V(y,z)$. Using the condition that
$V(y,z)=V=constant$, we obtain an equation for $y$ of the form:
$F(y,\dot{y})=V$.  From this equation, we can calculate the
relaxation oscillation period ($T$) by integrating over a cycle.
This  leads to the expression:
\begin{equation}
\label{tv}
T=\int_{y_0}^{y_1} \frac{1+\bar{s}y}{y}
\frac{dy}{{[2 (V - A_1 y - A_2 y^2 - A_3 \ln(y) -A_4 y^{-1})]}^{1/2}}
\end{equation}
where $y_0$ and $y_1$ are the values of $y$ that cancel the
denominator. We stress the fact that the only one approximation used
in the derivation of this expression is that the Lyapunov potential
is constant during two maxima of the intensity oscillations. The
previous equation for the period reduces, in the case
$c=d=\bar{s}=0$, to the one previously obtained by using the relation
between the laser dynamics and the Toda oscillator derived
in~\cite{oppo}.  Evaluation of the above integral shows that the
period $T$ decreases as the potential $V$ decreases. Since the
Lyapunov potential decreases with time, this explains the fact that
the period of the oscillations in the transient regime decreases with time. In
Fig. 5 we compare the results obtained with the above expression for
the  period with the one obtained from numerical simulations of the
rate equations (\ref{ibar}), (\ref{x}). In the simulations we compute
the period as the time between two peaks in the evolution of the
variable $y$. As can be seen in this figure,  the above expression
for the period, when using the numerical value of the  potential $V$,
accurately reproduces the simulation results although it is
systematically lower than the numerical result. The discrepancy is
less than one percent over the whole range of times. 

It is possible to quantify the difference between the approximate
expression (\ref{tv}) and the exact values near the stationary state.
In this case expression (\ref{tv}) reduces to :
\begin{equation}
T=\frac{2 ~\pi}{d_{12,st}~\sqrt{E F - H^2}}
\label{peestap}
\end{equation}
where:
\begin{eqnarray}\nonumber
E & = & 2 \left( A_2-\frac{1}{2}\frac{A_3}{y_{st}^2}+\frac{A_4}{y_{st}^3}+
\frac{1}{2}\left[\bar{s}+\frac{(d+ c z_{st})}{2 y_{st}^2}\right]^2\right)
\\\nonumber
F & = & \left[ 1+c \frac{(1+\bar{s} y_{st})}{2 y_{st}}\right]^2\\\nonumber
H & = & -\left[1+\frac{c (1+\bar{s} y_{st})}{2 y_{st}}\right]
\left[\bar{s}+\frac{(d+c z_{st})}{2 y_{st}^2}\right]
\end{eqnarray}
and $d_{12,st}$ is the coefficient $d_{12}$ calculated in the steady
state. The period of the relaxation oscillations near the steady
state can be obtained by linearizing eqs. (\ref{ibar}) and (\ref{x})
after a small perturbation is applied. The frequency of the
oscillations in the steady state is the imaginary part of the
eigenvalues of the linearized  equations. This yields a period:
\begin{equation}
T_{st}=\frac{2  ~\pi}{d_{12,st}~\sqrt{E F - H^2}} 
\left[1 - \frac {d_{22,st}^2}{d_{12,st}^2} 
\frac {F^2}{4 ( E F - H^2 )} \right]^{-1/2}
\label{peest}
\end{equation} 

The difference between (\ref{peestap}) and (\ref{peest}) vanishes
with $d_{22,st}$ (i.e. $d_{22}$ in the stationary state). Since $E F
-H^2$ is always a positive quantity, our approximation will give, at
least asymptotically, a smaller value for the period.

In order to have a complete understanding of the variation of the
period with time, we need to compute the time variation of the
potential $V(\tau)$ between two consecutive intensity peaks. This
variation  is induced by the dissipative terms in the equations of
motion. We have not been able to derive an expression for the
variation of the  potential (see \cite{oppo} for an approximate
expression in a simpler case). However, we have found that a
semi-empirical argument can yield a very simple law which is well
reproduced by the simulations. We start by studying the decay to the
stationary state in the linearized equations. By expanding around the
steady state: $y=y_{st}+\delta y$, $z=z_{st}+\delta z$, the dynamical
equations imply that the variables decay to the steady state as:
$\delta y (\tau)$, $\delta z(\tau) \propto \exp(-\frac{\rho}{2}\tau)$, 
where:
\begin{equation}
\rho=d_{22,st} F
\end{equation}
Expanding $V(y,z)$ around the steady state and taking an initial
condition at  $\tau_{0}$ we find an expression for the decay of the
potential:
\begin{equation}
\ln~[V(\tau)-V_{st}]=\ln~[V(\tau_{0})-V_{st}]-\rho~(\tau-\tau_{0})
\label{lnvtauaj}
\end{equation}
In Fig. 6 we plot $\ln[V(\tau)-V_{st}]$ versus time 
and compare it with the approximation (\ref{lnvtauaj}).
One can see that the latter fits $\ln[V(\tau)-V_{st}]$
not only near the steady state (where it was derived),
but also during the transient dynamics.
The value of $\tau_{0}$, being a free parameter, was chosen at the
time at which the first peak of the intensity appears. Although other
values of $\tau_{0}$ might produce a better fit, the one chosen here
has the advantage that it can be calculated analytically by following
the technique of ref. \cite{balle}.
It can be derived from Eq. (\ref{tv}) that the period $T$ is linearly 
related to
the potential $V$. This, combined with the
result of Eq. (\ref{lnvtauaj}), suggests the semi-empirical law for 
the evolution of the period of the form:
\begin{equation}
\ln~[T(\tau)-T_{st}]=\ln~[T(\tau_{0})-T_{st}]-\rho~(\tau-\tau_{0})
\label{lnTtauaj}
\end{equation}
This simple expression fits well the calculated period not only near
the steady state, but also in  the transient regime, see Figs. 5 and 7. 
The tiny differences observed near the steady state are due to 
   the fact that the semi-empirical law, Eq. (\ref{lnTtauaj}), is based on the validity 
   of relation  Eq. (\ref{tv}) between the period and the potential. As it was already
   discussed above, that
   expansion slightly underestimates the asymptotic 
   (stationary) value  of the period.
By complementing this study with the procedure 
given in \cite{balle} to describe the switch-on process of a laser,
and valid until the first intensity peak is reached, we can obtain a
complete description of the variation of the oscillations period in
the dynamical evolution following the laser switch-on.

\section{Summary}
\label{summary}
In this work we have used Lyapunov potentials in the context of laser
dynamics. For class A lasers, we have explained qualitatively the
observed features of the deterministic dynamics by the movement on
the potential landscape. We have identified the relaxational and
conservative terms in the dynamical equations of motion. In the
stochastic dynamics (when additive noise is added to the equations),
we have explained the presence of a ``noise sustained flow" for the
phase of the electric field as the interaction of  the conservative
terms with the noise terms. An analytical expression allows the
calculation of the phase drift.

In the case of class B lasers, we have obtained a Lyapunov potential
valid only in the deterministic case, when noise fluctuations are
neglected. We have found that the dynamics is non-relaxational with a
non-constant matrix $D$.  The fixed point corresponding to the laser
in the {\sl on} state is interpreted as a minimum in the potential
landscape. By observing that the potential is nearly constant between
two consecutive intensity peaks during the relaxation process towards
the steady state, but still in a highly non-linear regime, we were
able to obtain an approximate expression for the period of the
oscillations. Moreover, we have derived a simple exponential approach
of the period of the oscillations with time towards the period of the
relaxation oscillations near the steady state. This  dependence
appears to be valid after the first intensity peak following  the
switch-on of the laser.  A possible extension of our work could be to
consider the presence of an external  field, which is numerically
studied in \cite{oppo2}.

\section*{Acknowledgments}
We wish to thank Professor M. San Miguel and Professor G.L. Oppo for a
careful reading of this manuscript and for useful comments. We acknowledge
financial support from DGES (Spain) under Project Nos. PB94-1167 and PB97-0141-C02-01.

\end{multicols}

\vspace{8.0cm}
\begin{center}
Table 1
\vspace{2.0cm}

\begin{tabular}{|c|l|c|}\hline
\label{tabla1}
PARAMETERS & & VALUES\\\hline
$C$ & Carriers injected per unit time. & $> {\rm threshold}$\\ \hline
$\gamma$ & Cavity decay rate. & $0.5 ~ps^{-1}$ \\\hline
$\gamma_e$ & Carrier decay rate. & $0.001~ps^{-1}$ \\ \hline
$N_o$ & Number of carriers at transparency. & $1.5 \times 10^8$ \\\hline
$g$ & Differential gain parameter. & $1.5 \times 10^{-8}~ps^{-1}$ \\\hline
$s$ & Saturation parameter. &  $10^{-8}-10^{-7}$ \\\hline
$\beta$ & Spontaneous emission rate. & $10^{-8} ps^{-1}$ \\\hline
$\alpha$ & Linewidth enhancement factor. & 3-6 \\\hline
\end{tabular}
\end{center}

\newpage

\section*{Figure captions}

\vfill 
\noindent Fig. 1: Potential for a class A laser, Eq. (\ref{potclasa})
with the parameters: $a=2$, $b=1$. Dimensionless units.
\vfill 
\noindent Fig. 2: Time evolution of the mean value of  the phase
$\phi$ in a class A laser, in the case $a=2$, $b=1$, $\epsilon=0.1$. 
For $\alpha=0$ (dashed line)
there is only phase diffusion and the average value is $0$ for all
times. When $\alpha=5$ (solid line) there is a linear variation of the
mean value of the phase at late times. Error bars are incluted
for some values. The dot-dashed line has the
slope given by the theoretical prediction Eq. (\ref{fist}). The
initial condition is taken as $x_1=x_2=0$ and the results were 
averaged over 10000 trajectories with different realizations of the
noise. Dimensionless units.
\vfill
\noindent Fig. 3: a) Normalized intensity,  $y$ (solid line) and
normalized carriers number, $z/40$ (dot-dashed line) versus time in a
class B laser obtained by numerical solution of Eqs. (\ref{ibar}) and
(\ref{x}). b) Plot of the potential (\ref{vcbs}). Parameters:
$a=0.009$, $b=0.004$, $\bar{s}=0.5$, $c=3.2 \times 10^{-9}$, 
$d=1.44\times 10^{-8}$ which correspond to physical parameters in
Table 1 with $C=1.2 C_{th}$. The initial conditions are taken as
$y=5 \times 10^{-8}$ and $z=0.993$. Dimensionless units.
\vfill
\noindent Fig. 4: Number of carriers versus intensity (scaled
variables). The vector field and contour plot (thick lines) are also represented. 
 Same parameters
than in Fig. 3. Dimensionless units.
\vfill 
\noindent Fig. 5: Period versus time in a class B laser. Solid line
has been  calculated as the distance between two peaks of intensity,
with triangles  plotted at the begining of each period; dashed line
has been calculated using the expression (\ref{tv}), with the value
of the potential $V$ obtained also from the simulation; dotted line
corresponds to the semi-empirical expression (\ref{lnTtauaj}). Same
parameters than in Fig. 3. We have used $\tau_0=55.55$,
coinciding with the position of the first intensity peak. Dimensionless units.
\vfill
\noindent Fig. 6: Logarithm of the  potential difference versus time
in a class B laser (solid line), compared with the theoretical
expression in the steady state (\ref{lnvtauaj}) (dashed line).  Same
parameters than in Fig. 3 and $\tau_0$ as in Fig. 5. Dimensionless units. 
\vfill
\noindent Fig. 7: Logarithm of the period  difference versus time in
a class B laser. Triangles correspond to the period calculated from
the simulations as the distance between two consecutive intensity
peaks, at the same position than in Fig. 6. The dashed line is the
semiempirical expression Eq. (\ref{lnTtauaj}). Same parameters than
in Fig. 3 and $\tau_0$ as in Fig. 5. Dimensionless units.

\newpage
\centerline{\psfig{figure=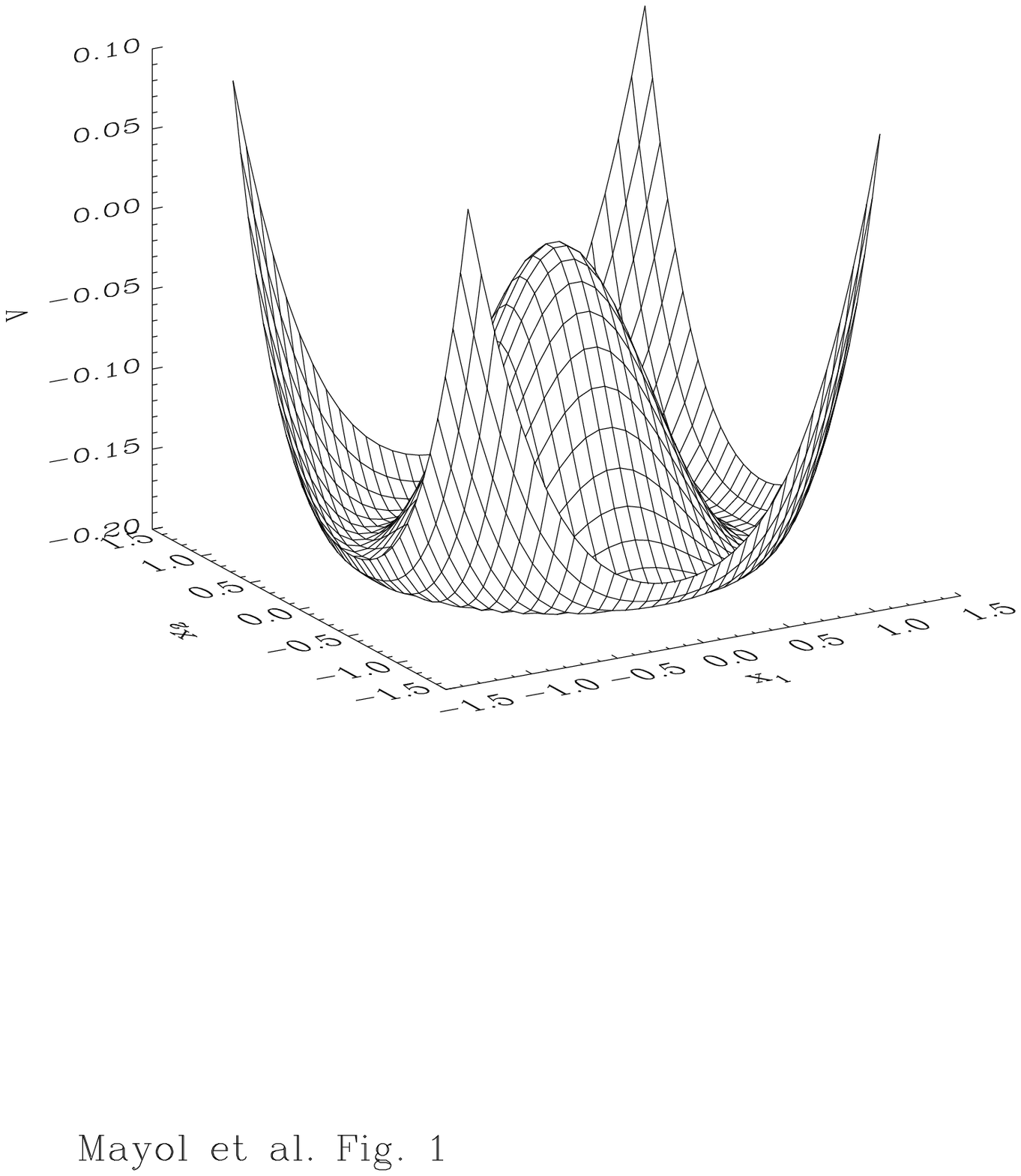,width=9.5cm,height=7cm}}
\vspace{5.0truecm}
\centerline{\psfig{figure=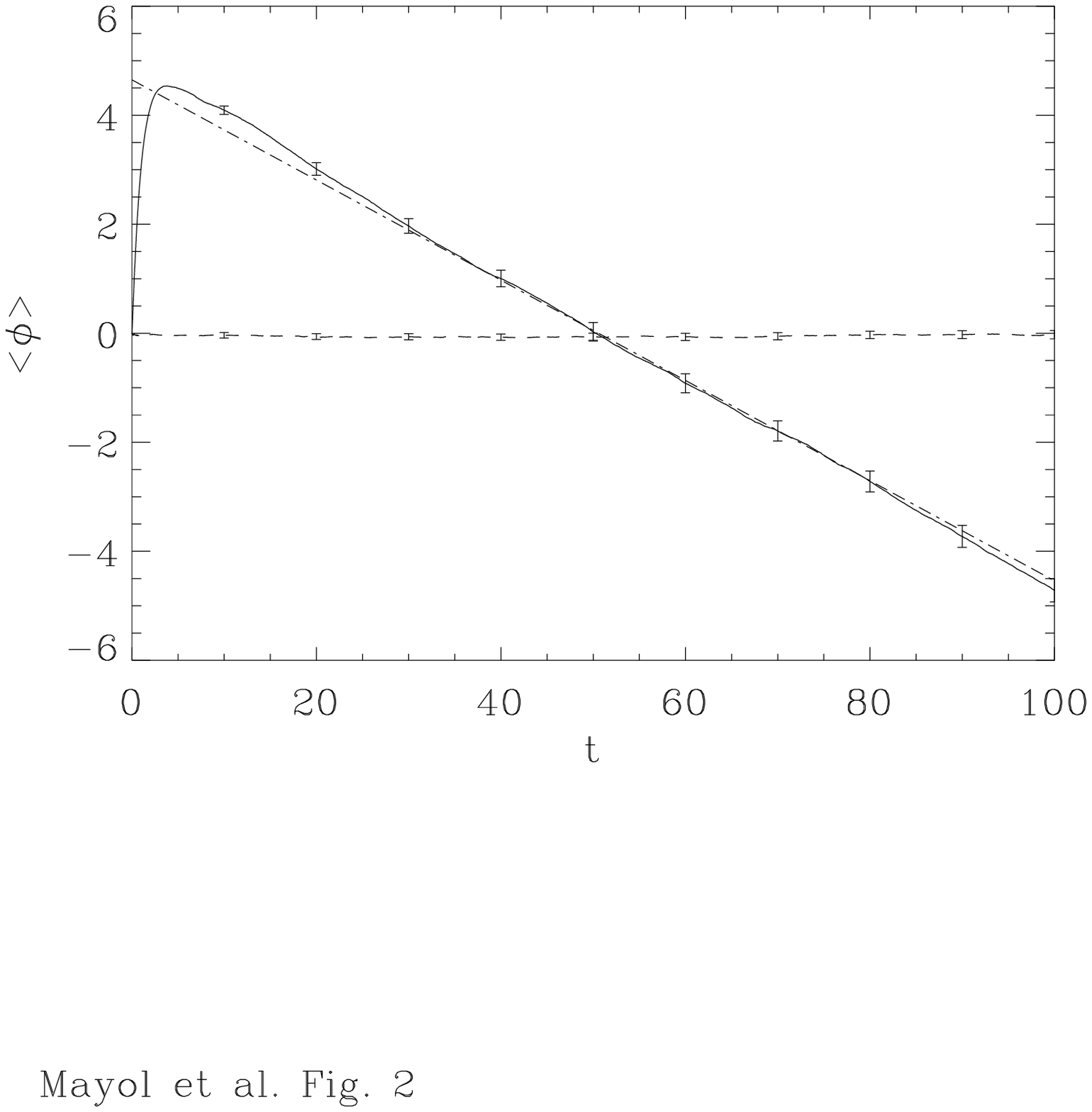,width=9.5cm,height=7cm}}
\newpage
\centerline{\psfig{figure=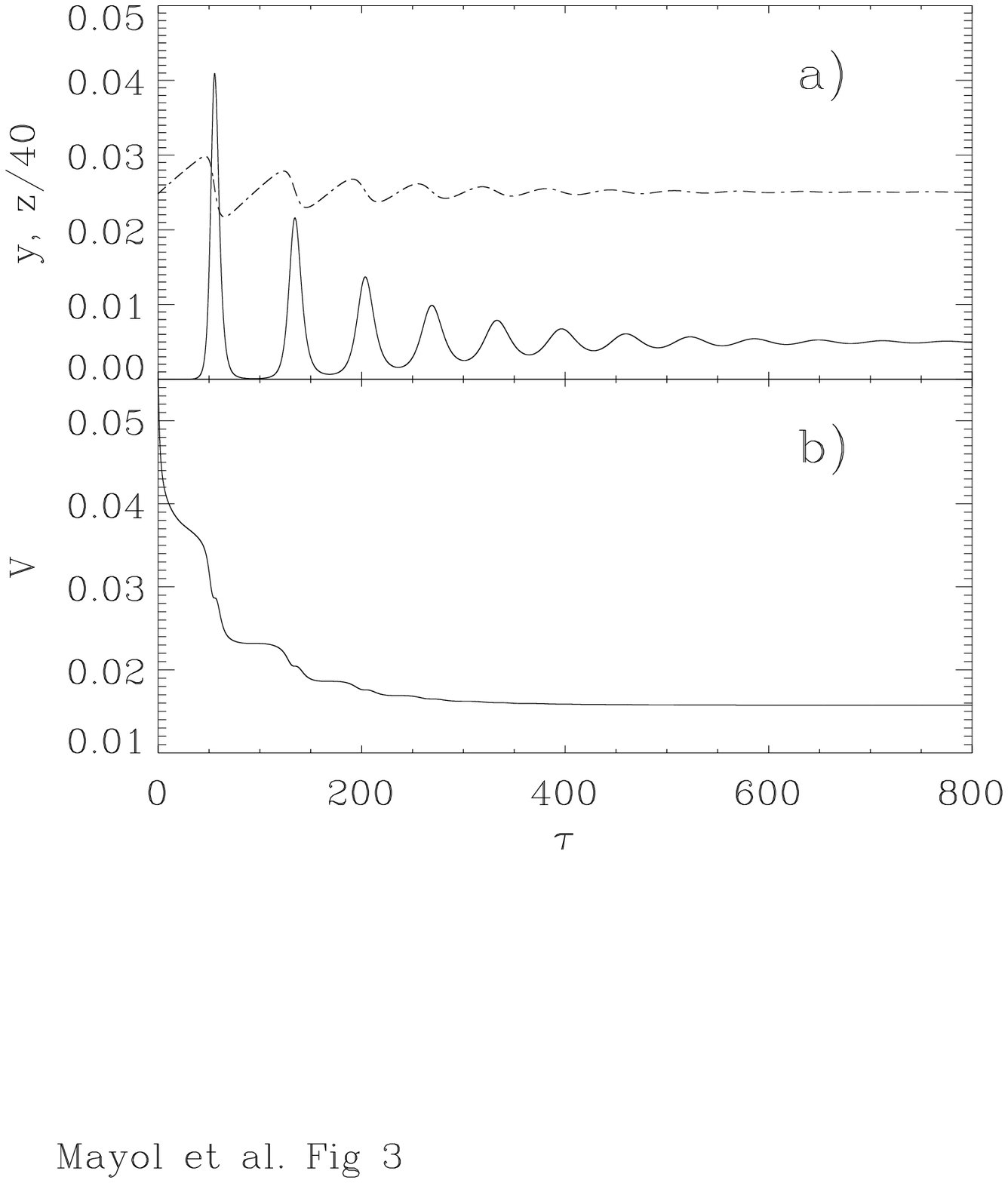,width=9.5cm,height=7cm}}
\vspace{5.0truecm}
\centerline{\psfig{figure=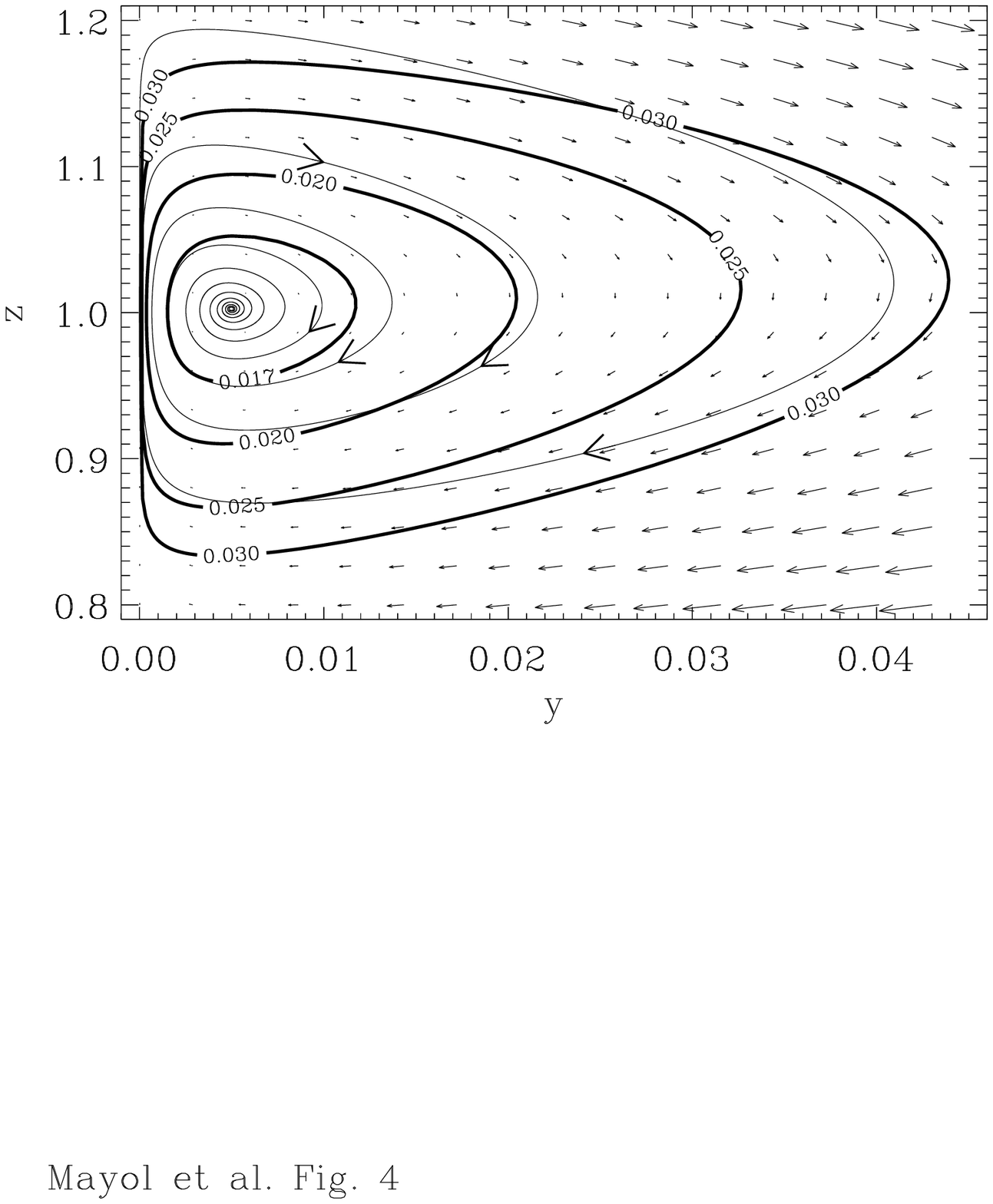,width=9.5cm,height=7cm}}
\newpage
\centerline{\psfig{figure=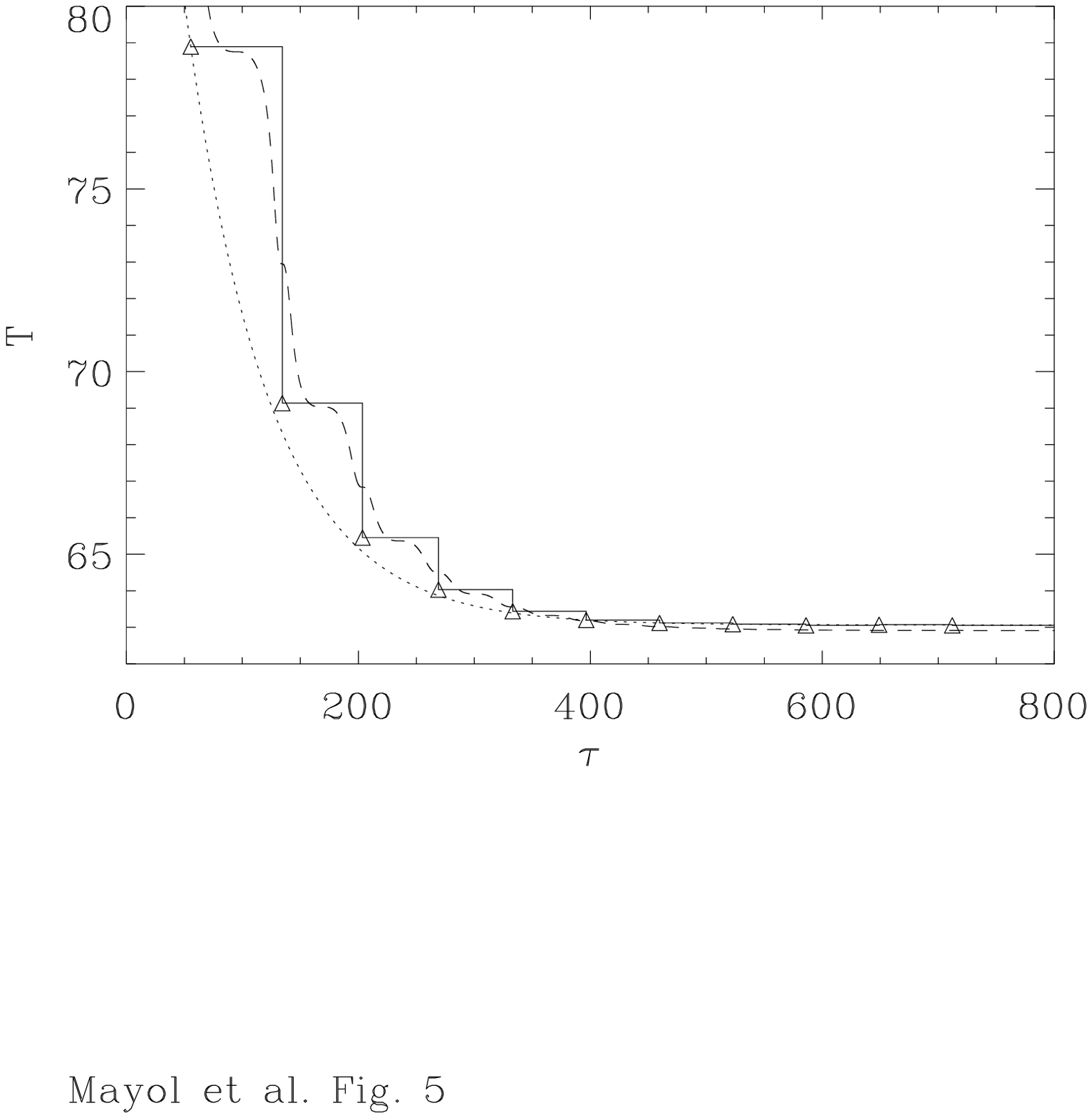,width=9.5cm,height=7cm}}
\vspace{5.0truecm}
\centerline{\psfig{figure=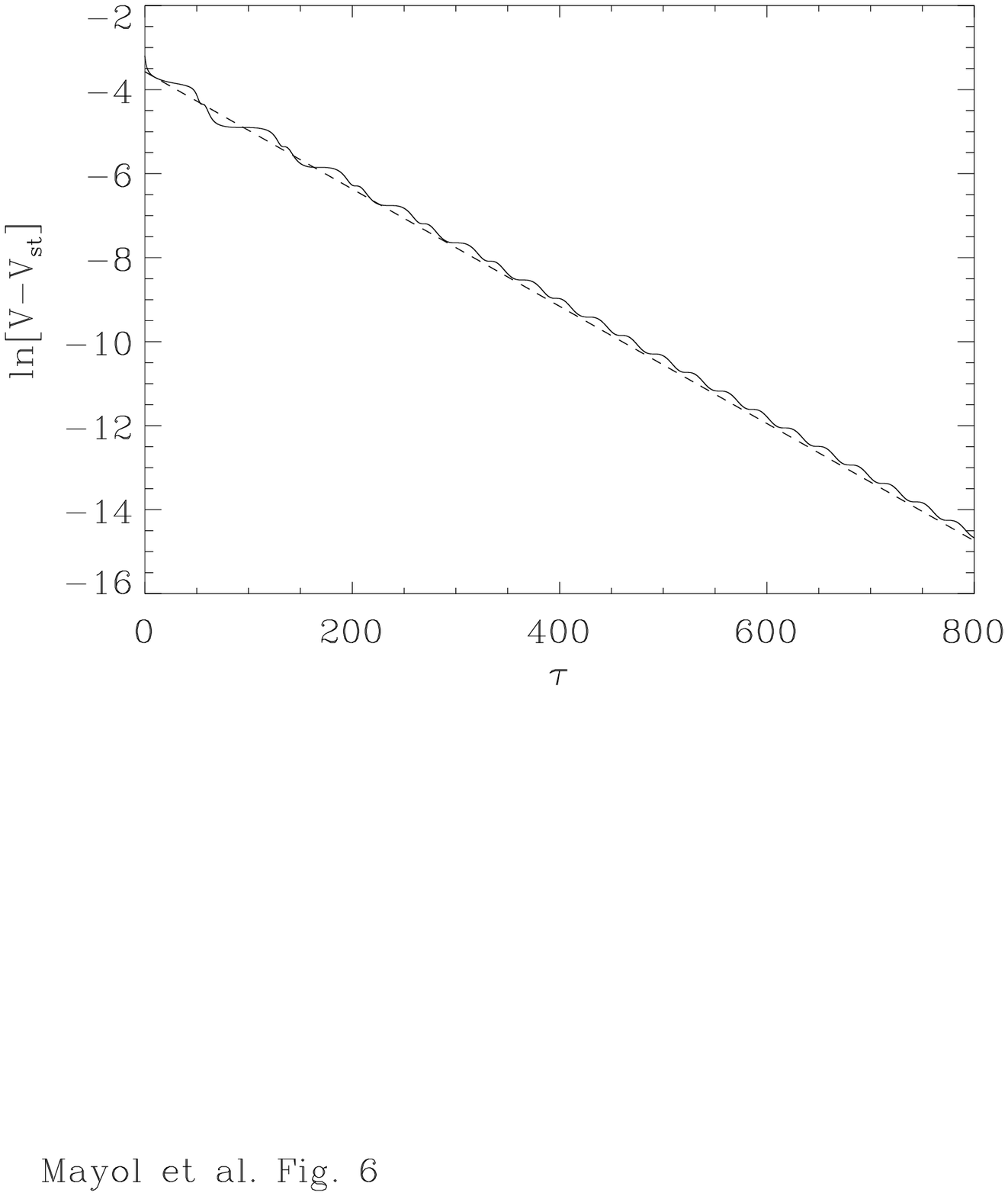,width=9.5cm,height=7cm}}
\newpage
\centerline{\psfig{figure=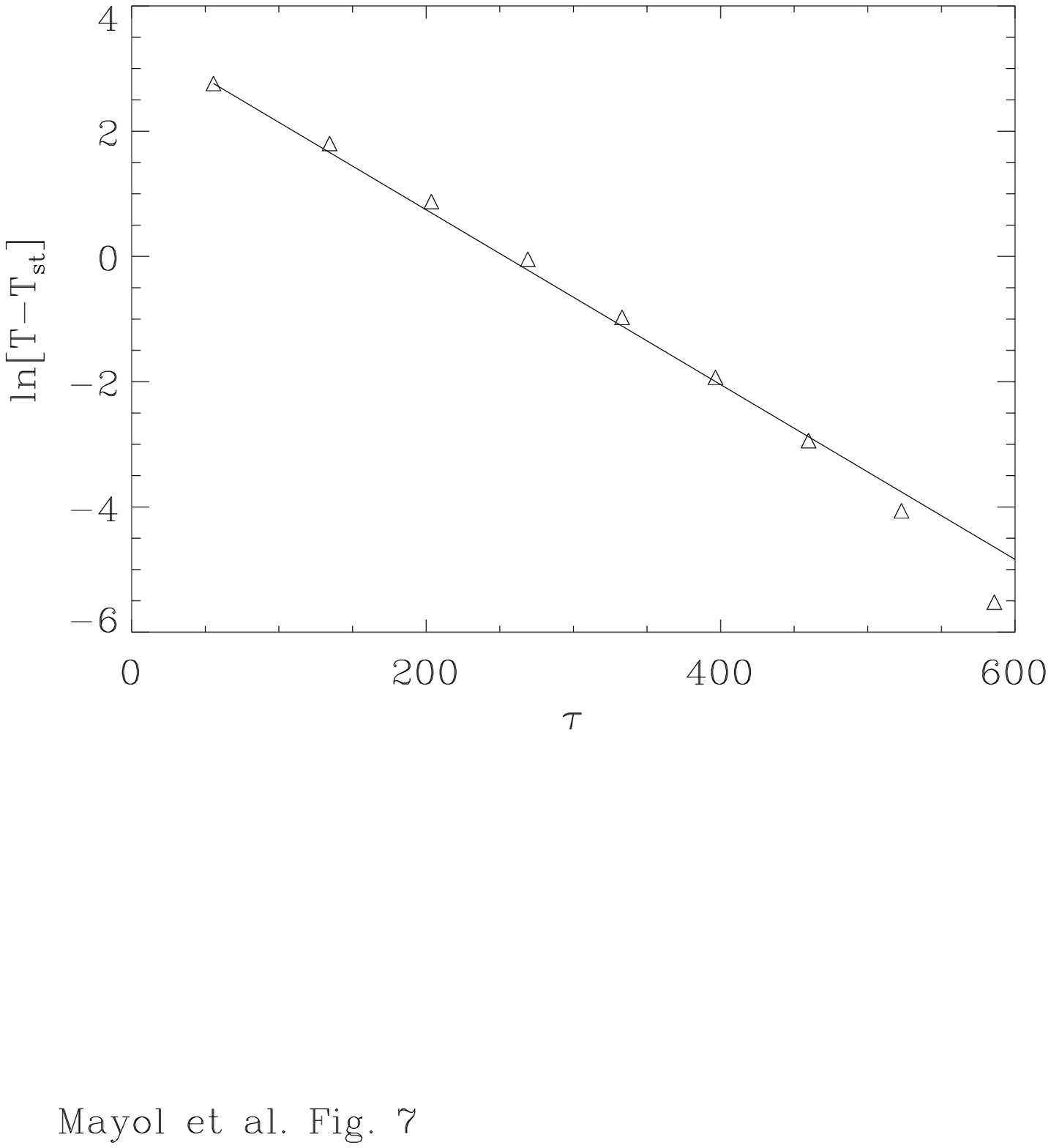,width=9.5cm,height=7cm}}
\end{document}